# A High-Throughput and Data-Driven Computational Framework for Novel Quantum Materials

Srihari M. Kastuar,[1, a)] Christopher Rzepa,[2] Srinivas Rangarajan,[2] and Chinedu E. Ekuma[1, b)]
[1)]*Department of Physics, Lehigh University, Bethlehem, PA 18015, USA*
[2)]*Department of Chemical and Biomolecular Engineering, Lehigh University, Bethlehem, PA 18015, USA*

Two-dimensional layered materials, such as transition metal dichalcogenides (TMDs), possess intrinsic van der Waals gap at the layer interface allowing for remarkable tunability of the optoelectronic features via external intercalation of foreign guests such as atoms, ions, or molecules. Herein, we introduce a high-throughput, data-driven computational framework for the design of novel quantum materials derived from intercalating planar conjugated organic molecules into bilayer transition metal dichalcogenides and dioxides. By combining first-principles methods, material informatics, and machine learning, we characterize the energetic and mechanical stability of this new class of materials and identify the fifty (50) most stable hybrid materials from a vast configurational space comprising $\sim 10^5$ materials, employing intercalation energy as the screening criterion.

## I. Introduction

Two-dimensional (2D) transition metal dichalcogenides (TMDs) are distinguished within the realm of 2D materials for their remarkable adaptability. Their layered crystal structure, marked by inherent interlayer van der Waals (vdW) forces and transition metals with partially filled d orbitals, endows 2D TMDs with an array of intriguing behaviors. These include layer-dependent electronic, optoelectronic, mechanical, and thermal properties, making 2D TMDs prime candidates for advanced applications in flexible electronics, photovoltaics, and biomedical technologies.[1–5] Furthermore, the optoelectronic characteristics of 2D TMDs can be fine-tuned by manipulating the crystal structure and the elemental makeup.[6–8] A particularly effective method for modulating properties in these materials involves the intercalation of foreign entities, such as metals or molecules, within the intrinsic vdW gap.[9,10] Such intercalation can modify the van der Waals bond strength and the composition of the host material, affecting doping levels, charge carrier density, mobility, and introducing novel energy states.[11–14] Importantly, the intercalation process in 2D TMDs is reversible, which preserves the lattice structure and allows for precise control of the properties by varying the type and concentration of the intercalants. Organic molecules, such as conjugated molecules, have proven instrumental in creating low-cost, flexible devices.[15–19] Furthermore, previous studies highlight improved device performance when organic semiconductors are integrated into 2D materials;[20,21] these make them an ideal candidate for intercalating into 2D TMDs to tune and improve their optoelectronic properties. As an example, we illustrate in Figure 1 the tunability of the energy bandgap in bilayer $MoS_2$ due to the intercalation of diverse conjugated organic molecules.

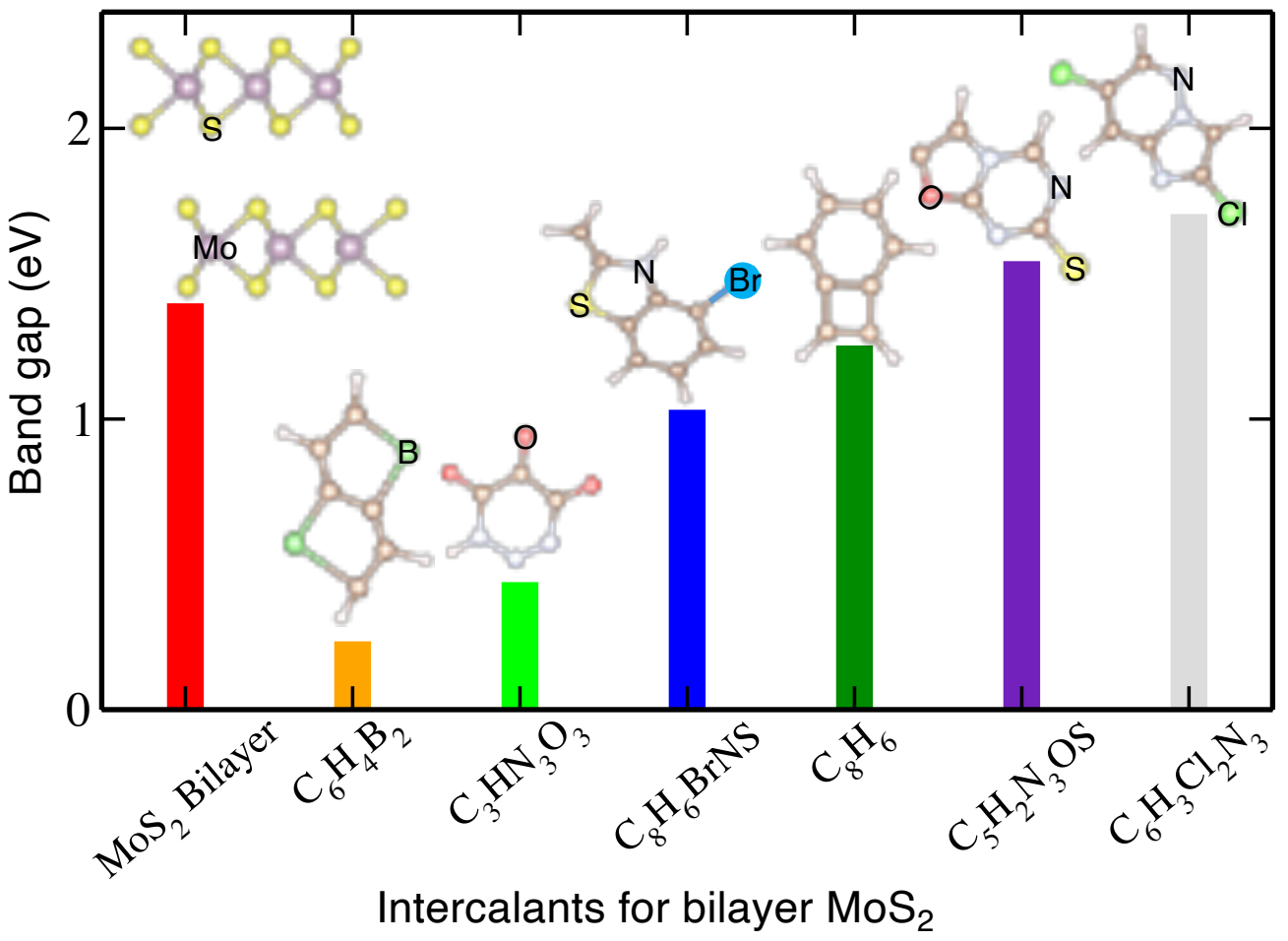

FIG. 1. The bandgap of pristine and organic molecule-intercalated bilayer $MoS_2$ was calculated using the PBE exchange-correlation functional within density functional theory.

To enable efficient exploration of the vast compositional space, we have developed a robust computational framework that employs active learning (AL), anchored in first-principles calculations and enhanced by materials informatics. This innovative strategy is applied to the design of new materials by intercalating conjugated organic molecules within the intrinsic van der Waals gap of bilayer transition metal dichalcogenides and dioxides, creating a diverse array of $MX_2$/organic hybrids, where M = Mo, W, Cr, Hf, Ti, Zr, and X = S, Se, Te, O. Using our method, we have successfully generated and scrutinized a vast compositional space of nearly 300,000 potential $MX_2$/organic hybrids, identifying those with the most favorable energetic and mechanical stability. This strategic approach holds the potential for broadening to include intercalated bilayer heterostructures and a variety of molecular intercalants, paving the way for the

[a)]Electronic mail: smk520@lehigh.edu
[b)]Electronic mail: che218@lehigh.edu

discovery of advanced materials with tailored properties.

## II. Method

### A. DFT Computational Details

The database for $MX_2$/organic hybrid materials was created using the 2H crystal symmetry of atomically thin $MX_2$ bilayers and conjugated organic molecular intercalants. Hybrid structures were modeled on a $3 \times 3 \times 1$ supercell of a bilayer exfoliated from bulk $MX_2$, employing first-principles density functional theory (DFT)[22,23] as implemented in the Vienna *ab initio* simulation package (VASP)[24,25]. The structural optimization of the parent materials and the corresponding high-throughput computations to assess the structural and energetic stability of the hybrid materials utilized the generalized gradient approximation (GGA) with the Perdew-Burke-Ernzerhof (PBE)[26] exchange-correlation functional. To ensure accurate initial configurations, the bulk $MX_2$ structures were calculated using a plane wave energy cutoff of 500 eV and an $8 \times 8 \times 8$ $\Gamma$-centered mesh to sample the Brillouin zone. For high-throughput calculations, $\Gamma$ point sampling of the Brillouin zone was used with a cutoff energy of 400 eV with the total energy and residual interatomic forces converging to $\sim 10^{-4}$ eV and 0.02 eV/Å, respectively. Despite the potential increases in absolute error from lower energy cutoff and k-space sampling in DFT calculations exceeding our target accuracy, the intercalation energy calculations benefit from significant error cancelation. This benefit arises because intercalation energy is determined by the energy differences between two similar states of the host material—before and after intercalation. These states typically share similar atomic densities and average coordination numbers, i.e., systematic errors from DFT approximations such as reduced energy cutoffs or limited k-space sampling, which similarly impact both the total energy calculations of the states. This structural and electronic similarity ensures consistent overestimations or underestimations in energy across both states due to DFT approximations. Consequently, when calculating the energy difference to determine intercalation energy, many of these errors cancel out. This error cancelation is especially advantageous, allowing for fast and accurate intercalation energy estimations and minimizing inherent computational challenges such as computational errors in high-throughput calculations, even when the absolute DFT energies of the individual states may not achieve the desired accuracy due to computational simplifications. In our study, we evaluated our selected input parameters by comparing them with stricter inputs (500 eV cutoff energy and a $2 \times 2 \times 1$ k-points mesh) across randomly chosen hybrid materials. We observed small absolute errors ranging from 0.1 to 0.8 eV in the intercalation energy, but the coarse input requires only one-sixth of the computational cost. To model the electronic features of the host materials and the top materials, we employed the Heyd-Scuseria-Ernzerhof (HSE06) hybrid functional[27] with a Hartree-Fock mixing parameter of 33%. This mixing ratio has been shown to better describe the electronic properties of 2D materials.[9,28,29] All calculations included van der Waals interactions using the DFT-D3 parameterization.[30]

### B. Material Design Pipeline

The creation of the initial crystal structures for the hybrid materials was accomplished via a tailored Python pipeline[31]. This pipeline comprises four principal stages, each crucial in creating an optimized structural model. In the first stage, the conjugated organic molecules were relaxed in an isolated gas phase. This phase was essential to obtain the most stable molecular configurations without the influence of any external intermolecular forces or potentials, thereby ensuring an accurate representation of the molecular structure. The second stage entailed the structural optimization and relaxation of the bulk $MX_2$ structures. After achieving an optimal configuration, the $MX_2$ layers were exfoliated to form atomically thin bilayers. This stage ensured that the structures were at their lowest energy state and provided a realistic framework for intercalation in the next step. The third stage involved the intercalation of the optimized conjugated organic molecules into these optimized, atomically thin bilayers. The incorporation of organic molecules within the $MX_2$ host layers was carefully managed to avoid any overlap of the adsorbate atoms with those in the bilayer or the adjacent cells. Finally, in the fourth stage, a subsequent structural relaxation was performed on the intercalated hybrid systems. This relaxation process allowed the hybrid system to reach its minimum energy configuration, thereby providing the most stable and reliable models for future studies.

We used a subset of conjugated organic molecules from the PubChem database,[32] which hosts more than a million such molecules. To ensure the compatibility of our chosen molecules with our investigation and to guide synthesis feasibility, we carefully selected those that satisfied certain predefined criteria. The distance between two atoms in the molecule was restricted to be smaller than the maximum diagonal distance of the bilayer unit cell. This ensured that the molecules fit within the unit cell of the $MX_2$ bilayers. Our preference was for cyclic molecules, primarily because of their inherent planarity. Specifically, we selected those with a best-fit plane (PBF) value less than $10^{-4}$.[33] This strategic selection was designed to protect the structural integrity of the host $MX_2$ layers. Additionally, our selection was contingent upon the molecules being charge neutral, devoid of uncommon isotopes or valencies. Moreover, to maintain a manageable computational expense and facilitate practical chemical synthesis, we constrained the atom types within the molecules to B, C, N, O, F, P, S, Cl, Se, and Br. As a result of these constraints, we ended up with a pool of 7,745 molecules. Upon selection, these molecules were converted from their corresponding SMILES[34] strings into three-dimensional (3D) structures using the MMFF94s force field[35,36] within the RDkit package.[37] These 3D molecular models were enclosed in a box with a minimum vacuum of 10 Å between periodic images (Figure 2a), providing an appropriate initial estimate for the subsequent DFT relaxation process.

To intercalate the structures, we performed a series of calculations within our automation pipeline. Initially, a $3 \times 3 \times 1$ supercell of the bulk structure of 2H-$MX_2$ (M = Mo, W, Cr, Hf, Ti, Zr; X = S, Se, Te, O) crystal was constructed with the $P6_3/mmc$, space group no. 194 primitive cells[38]. The unit cell shape, volume, and atom positions were simultaneously relaxed. After achieving convergence, two consecutive M-X layers were cleaved from the bulk by introducing a vacuum of 15 Å along the crystallographic c-axis. This process resulted in 24 $MX_2$-based bilayer structures with a crystal symmetry of $P\bar{3}m1$, space group no. 164 (see Figure 2b). The atomic positions within the bilayer were then relaxed and the organic molecules were automatically inserted into the natural van der Waals gap between the bilayer structures. The gap was further expanded by 3 Å to accommodate the molecules, and their PBF was oriented parallel to the bilayer surface. In our material design process, we aligned the maximum atom-to-atom distance in the organic molecule with the longer of the two diagonals in the unit cell. This approach ensured a minimum non-hydrogen atom-to-atom distance of 3.25 Å between periodic images, with no observed interactions between these images. To aid in the relaxation process of the hybrid structures, we kept the bottom layer structure of the host frozen.

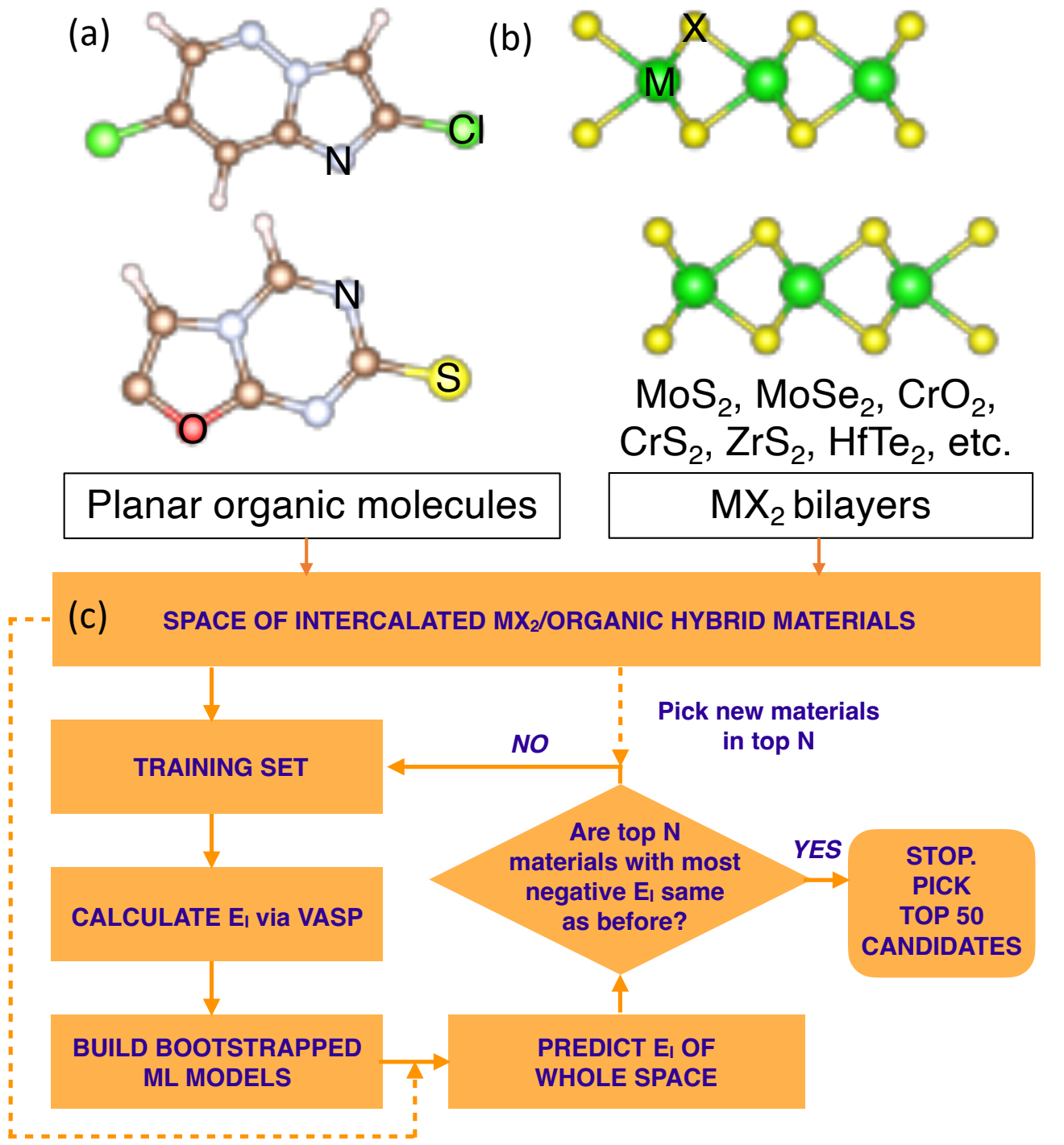

FIG. 2. (a) The space of 7,745 planar organic molecules extracted from PubChem. (b) The space of 24 2D bilayer transition metal dichalcogenides and dioxides of the form $MX_2$, e.g, $CrO_2$, $MoO_2$, $TiS_2$, $ZrS_2$, etc.(c) A flowchart of the active learning algorithm to predict the 50 most stable materials formed by intercalating a 2D bilayer with an organic molecule.

### C. Data-driven Strategy

Machine learning and material informatics approaches have emerged as powerful tools for inferring property-structure relations in vast configurational spaces, thereby accelerating the design of novel materials. Given the generation of roughly 300,000 hybrid structures from our materials design pipeline, employing data-driven approaches is crucial to identify the most promising structures. A key aspect of constructing such machine-learned models lies in the selection of features. We anticipate that a combination of structural features from the intercalant, such as the topological polar surface area, (TPSA), pathway fingerprints, and descriptors like the molecule's span and geometric diameter,[39] along with 2D material structural attributes such as the lattice constant and electronic fingerprints such as energetics, will suffice for model development.

The energetic features are essential for assessing the feasibility of synthesizing materials experimentally. In particular, the intercalation energy ($E_I$) – the energy needed to insert a guest molecule into a host material – effectively measures the stability and viability of intercalated structures. Although $E_I$ can be computed for any material, such computations for our entire configurational space are challenging. We expedited the material design by employing an active learning algorithm (Figure 2c) and created an ensemble of $\sim 10^2$ models by bootstrapping to assess the model uncertainty. Our model ensemble predicts the $E_I$ distribution across all potential materials, enabling us to rank them. By adopting an epsilon-greedy approach, we balance exploration and exploitation to determine candidates with the most negative $E_I$ – indicative of spontaneous intercalation – and those with significant uncertainty, characterized by their standard deviation. These candidates undergo further validation by DFT. We refined our machine learning model selection to increase accuracy by testing the performance of a variety of models ranging from simple regressors to advanced gradient boosting techniques. In particular, CatBoost[40] models surpass other machine learning approaches in performance (see Table S1 in the Supplementary Material (SM)). With new data, we update and reassess the models, continuously refining our ranking of the top materials and monitoring for deviations between successive iterations. If differences exceed our numerical uncertainty, we incorporate new materials from the plausible set into our training pool and repeat the process. If not, we stop the cycle and proceed to assess the mechanical stability of the leading materials. A detailed account of the material design pipeline is available in the SI. Our primary objective is to evaluate around $10^3$ candidates from a broader space, with the aim of a computational synthesis of ~50 materials.

## III. Results and Discussion

Our data-driven, high-throughput simulations typically require approximately five self-consistent active learning iterations to achieve convergence (Figure 3a). By the third iteration, around 41% of the top 50 materials had already been identified. In the fourth iteration, almost 100% of the top 50 materials were successfully identified, with the convergence of the algorithm confirmed in the fifth iteration. The convergence of the active learning process is directly related to the saturation of the performance of the machine learning models. The models were cross-validated using a training dataset that was updated with new materials in each iteration. The average $R^2$ score (mean squared error) of cross-validation tests decreased (increased) from 92% (0.04) in the second iteration to 76% (0.21) in the final iteration. This decline in performance metrics across iterations aligns with the exploitation phase, which promotes convergence as the model ceases to detect significant changes in the training data (see Figure S1).

Furthermore, in our approach, the uncertainty in predicting $E_I$ is consistently reduced due to the simultaneous *exploration* and *exploitation* of the configurational space, allowing new insights into uncertain materials (Figure 3b). With this balance, the 50 most stable materials were predicted with an average uncertainty of $\pm 0.32$ eV, with 75% of these materials having uncertainties less than $\pm 0.44$ eV. Verified by DFT computations, we report a root squared error (RMSE) of $\sim 0.6$ eV in these predictions, indicating a general tendency of the model to underestimate the strength of intercalation (Figure 3c). The reduction in uncertainty, combined with the observation that the predicted energies tend to under-bind the hybrid materials, ensures a sufficiently reliable prediction of stable materials while minimizing false-positive results (see SI for the stability analysis across iterations).

Our framework currently utilizes the CatBoost model; however, we believe that the performance of the active learning process can be significantly enhanced by incorporating advanced state-of-the-art architectures, such as graph neural networks (GNNs). These models are capable of capturing and integrating the inherent nonlinearity present in materials science datasets. Such sophisticated methods are crucial for more complex tasks, such as predicting optoelectronic properties (e.g., bandgap), where identifying the *best* material with minimal prediction errors is necessary for practical applications. However, training multiple GNNs with bootstrapped data is computationally expensive. Another approach to improve prediction quality is to increase *exploration* by randomly sampling additional data that are not directly related to the target property in successive iterations. This strategy may result in slower convergence of the active learning process. Nonetheless, with the rapid development of GNNs for materials science, and advancements in GPU architecture, our active learning framework holds the potential to describe more complex material properties, such as bandgap, exciton spectra, and beyond. These avenues will be explored in future work.

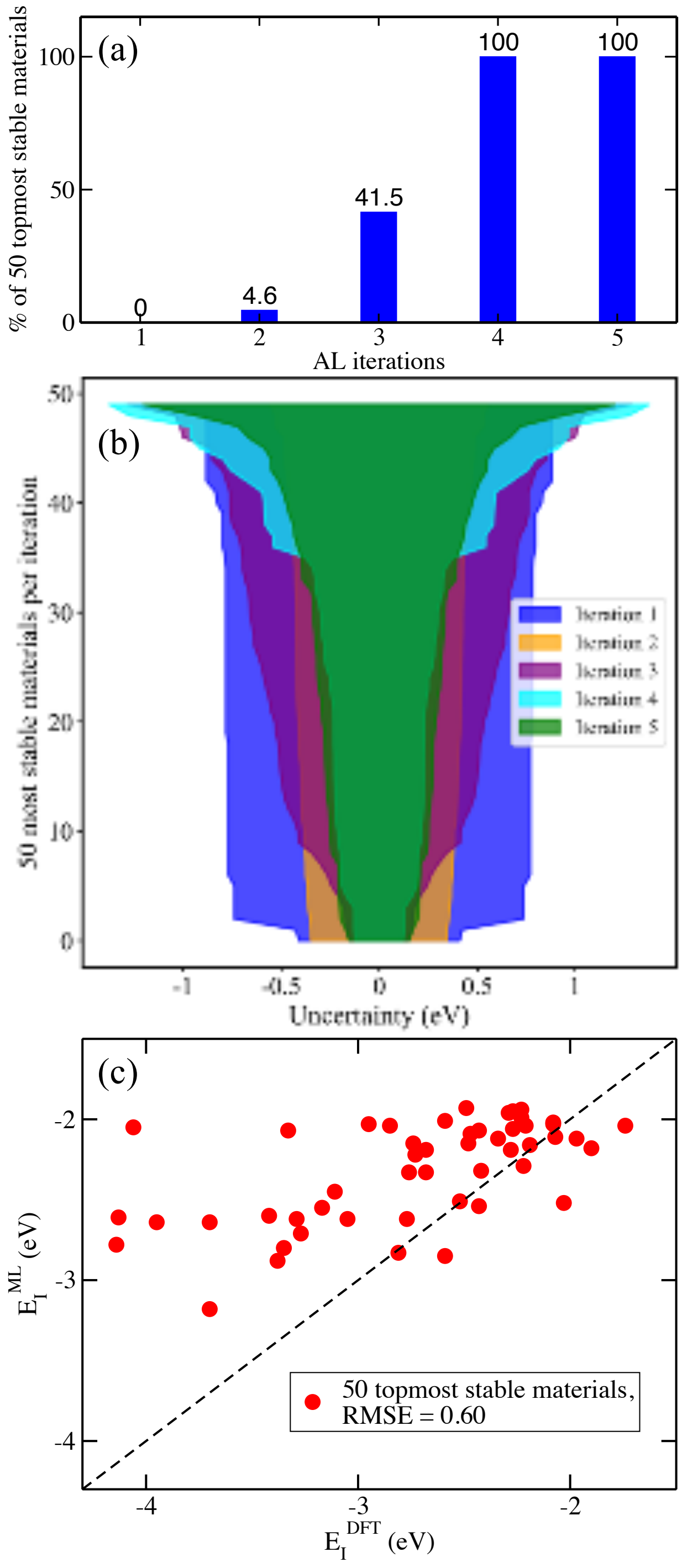

FIG. 3. (a) The fractional percentage of the 50 most stable hybrid materials represented in the top 150 stable hybrids per active learning (AL) iteration. (b) The evolution of the uncertainty band across the five AL iterations. The uncertainty band is formed by the standard deviation in predicted intercalation energies, with the mean shifted to zero. (c) The parity plot of the predicted ($E_I^{ML}$) versus computed ($E_I^{DFT}$) intercalation energies of the final 50 materials.

To further assess the feasibility of fabricating the predicted top 50 stable materials beyond their energetic stability, we explored their elastic and mechanical properties. These properties were calculated using the strain-stress method implemented in the ElasTool code[41–43]. The computed mechanical and elastic features are presented in Table S2 of the Supplementary Material. We observed that the elastic constants of the hybrid materials are approximately 10% higher than those of the parent bilayers, indicating enhanced mechanical robustness post-intercalation. The calculated elastic constants satisfy the Born stability criteria for hexagonal crystal structures[44,45], confirming that these materials possess the necessary mechanical stability. This criterion is crucial to predict the practical fabricability of these materials, as it rules out potential structural failures under mechanical stress. Additionally, Born's stability test captures low-energy phonon processes, providing a deeper understanding of the dynamic stability of the materials. Therefore, it serves as an essential second layer of screening in our high-throughput material design pipeline[43].

The predicted top 50 most stable $MX_2$/organic hybrid structures are dominated by organic intercalants containing N, S, O, F, and B. In particular, $CrO_2$-based hybrid materials represent 44 of the 50 top materials. Although pristine transition-metal dioxides are energetically more stable than dichalcogenides[46] - supported by their larger mechanical and elastic constants, which are more than double when compared to their oxide-based counterparts (see Table S3)—our calculations show that the intercalation of Hf, Zr, and Ti-based oxides significantly distorts the host lattice. In contrast, the unique electronic configuration of $CrO_2$ preserves its structural integrity during interactions with hydrogen-containing organic molecules. This stability is attributed to the unpaired 4s electron in Cr that allows for $CrO_2$ and its hybrids to form hydrogen-like bonds with minimal lattice distortion. The $CrO_2$-based hybrids were found to be more prone to spontaneous intercalation, leading to their dominance among the top 50 predicted hybrid materials from our active learning pipeline. By selectively filtering out $CrO_2$-based systems from the results of the final iteration, we curated a list of the most stable TMD-based hybrid structures. This refined list features a variety of highly sought-after TMDs, including $HfS_2$, $TiSe_2$, $TiTe_2$, and $ZrS_2$ as host structures. Our active learning approach, thus refined, predicts the intercalation energies of these topmost stable TMD-based materials with a mean absolute error of 0.45 eV and a RMSE of 0.56 eV (see Table S4), underscoring the model's robustness in accurately predicting the stability of a diverse range of $MX_2$/organic hybrid structures.

Intercalation is known to modify the properties of the host material through various effects. These include physical transformations such as the increase or decrease in interlayer spacing and chemical changes arising from the modified atomic composition. The reactivity of the intercalant can readily change the order of the interlayer bond from purely van der Waals interactions to a blend of van der Waals, covalent, and ionic interactions.[6] In the following section, we discuss the various interlayer interaction mechanisms and the corresponding changes in the vdW gap of the host materials post-intercalation. In these analyses, we have broadly categorized the interlayer interactions into purely van der Waals, van der Waals with a degree of covalency between the host and organic material (inter-species bonds) and van der Waals with hydrogen bonding between nearest-neighbor intercalants.

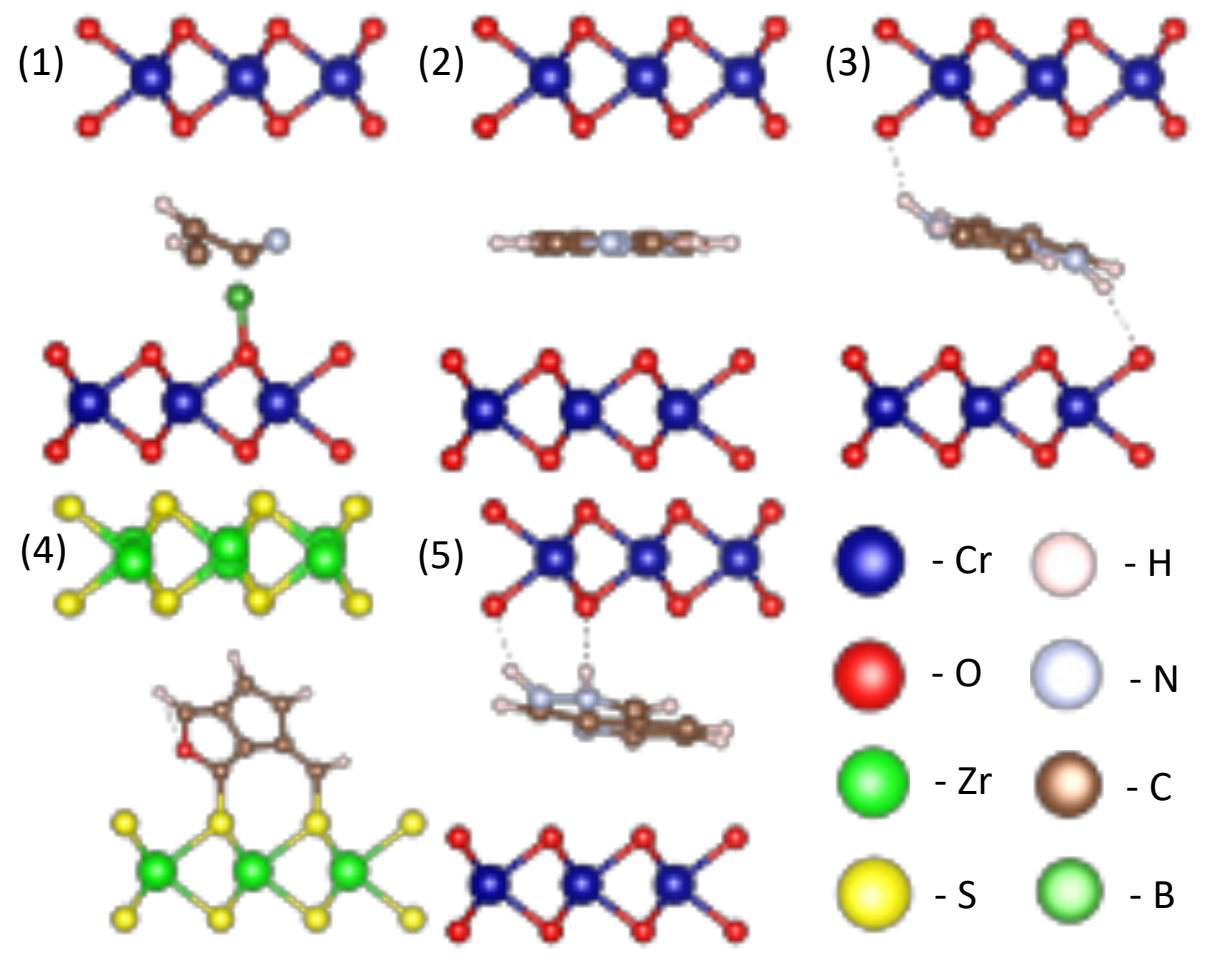

FIG. 4. The crystal structures of the five most stable $MX_2$/organic hybrid materials.

TABLE I. The top five stable 2D/organic hybrid materials based on the intercalation energy $E_I$ (eV), energy bandgap $E_g$ (eV), and change in the stiffness constant $\Delta K$ (N/m).

| Hybrid material | $E_I$ | $E_g$ | $\Delta K$ |
|---|---|---|---|
| $CrO_2$[B1=C(C=C1)F] | -4.14 | 0.40 | 23.47 |
| $CrO_2$[C1=CN=CC2=CNNC=C21] | -4.13 | 0.46 | 39.58 |
| $CrO_2$[C1=CNN2C=CNC2=C1] | -4.06 | 0.47 | 24.68 |
| $ZrS_2$[C1=CC2=COC3=CC1=C23] | -3.95 | 0.72 | 12.63 |
| $CrO_2$[C1=CC2=CNNC=C2N=C1] | -3.70 | 0.44 | 20.98 |

First, we re-rank our predicted 50 topmost hybrid materials on the basis of their DFT-calculated values of $E_I$. The structure of the five most stable hybrids according to the new ranking is presented in Figure 4. The corresponding $E_I$, electronic bandgap, and the change in the stiffness constant after the intercalation are presented in Table I. These quantities can be readily obtained from our high-throughput DFT calculations and are important for the effective screening of such a large material space. The top 5 hybrid materials contain one $ZrS_2$ and four $CrO_2$ hybrids. The large negative $E_I$ clearly points to their energetic stability. From our high-throughput computations at the PBE level of theory, we categorize the top five hybrids as small bandgap materials, with an average gap of 0.44 eV for $CrO_2$ and 0.72 eV for the $ZrS_2$ hybrid materials. We note that the top two materials, although very close in terms of energetic stability, show completely different bonding mechanisms. The $CrO_2$[B1=C(C=C1)F] hybrid is mostly dominated by inter-species B-O bonds of $\sim$ 1.40 Å, whereas in $CrO_2$[C1=CN=CC2=CNNC=C21] the intercalant remains suspended within the vdW gap with no visible interspecies bonds. In the third and fifth most stable hybrids, the intercalants were observed to bond with the $CrO_2$ layers with strong O-H bonds of $\sim$ 1.9 and 1.7 Å, respectively. The lower energetic stability of $CrO_2$[C1=CC2=CNNC=C2N=C1] can be attributed to the presence of two 6-member carbon rings that undergo considerable deformation after intercalation. While the intercalant in $CrO_2$[C1=CNN2C=CNC2=C1], made up of a six- and a five-member carbon ring, simply rotates off the x-y plane to bond with the host material. The fourth most stable hybrid was found to be $ZrS_2$[C1=CC2=COC3=CC1=C23]. Here, the intercalant consists of a four-member carbon ring that dissociates within the vdW gap, causing the molecule to rotate by $\sim 90°$ along the y-axis, leading to the formation of C-S bonds of 1.7 Å. We can further characterize the different bonding mechanisms using the mechanical fingerprints, such as the in-plane stiffness constant (K). Intuitively, the change in the in-plane stiffness constant should be related to the bonding mechanisms; specifically, the presence of hydrogen bonds between nearest-neighbor intercalants may lead to an increase in K. This behavior is partially observed in our analysis of the top 5 materials where the second most stable material shows the largest ΔK which we attribute to the anticipated hydrogen bonds between neighboring C1=CN=CC2=CNNC=C21 molecules. In general, in the 50 most stable materials, we note that the presence of hydrogen bonds between the nearest-neighbor intercalants leads to an average increase of 20%, while, in other cases, the average increase is about 8%.

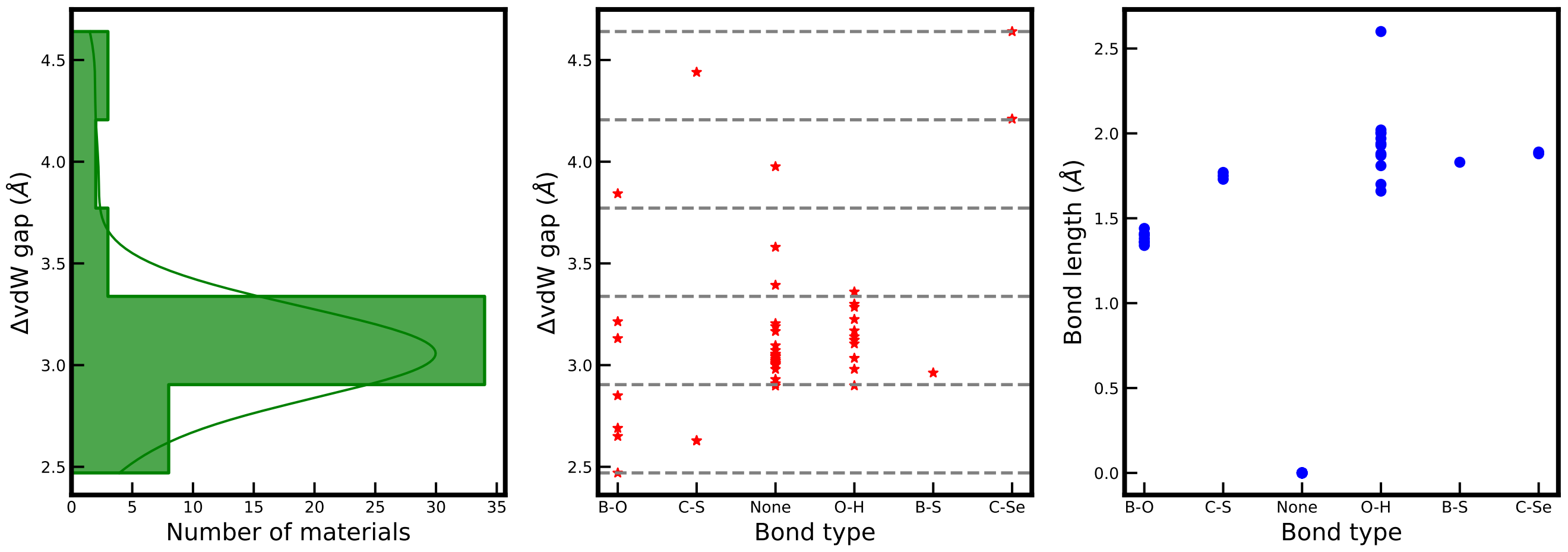

FIG. 5. The distribution of change in the van der Waals gap ΔvdW (left panel), the dominant interaction mechanism (middle panel), and the corresponding predicted bond lengths (right panel) in the top 50 hybrid materials.

Next, we analyze the change in the vdW gaps (ΔvdW) of the parent bilayers post-intercalation and the corresponding dominant interlayer interaction mechanism (Figure 5) in all the 50 topmost stable hybrid materials. We can group the bonding mechanisms between the host and the intercalant into six types: B-O, B-S, C-S, C-Se, O-H, and no bonds. The last category includes materials with bonds between neighboring intercalants; these bonds seem to have a negligible effect on the vdW gap. The average lengths of B-O, B-S, C-S, and C-Se bonds were found to be $\sim$ 1.4, 1.8, 1.7, and 1.9 Å, respectively. In the case of the O-H bonds, we can further divide them into weak and strong bonds with bond lengths of $\sim$ 1.7 and 2.7 Å, respectively. In the distribution of ΔvdW (left panel), the most frequent occurrences fall within the range of 2.8 to 3.3 Å. These materials primarily exhibit dominant interspecies interactions involving strong O-H bonds and purely van der Waals interactions (middle panel). Furthermore, substantial ΔvdW ($>$ 4 Å) occurs in exceptional cases where the intercalant undergoes significant structural rearrangement. For example, C-X (X = S, Se) bonds formed in $ZrX_2$[C1=CC2=COC3=CC1=C23] hybrids result in an approximate 4.5 Å increase in the van der Waals gap. Intriguingly, the formation of B-O bonds appears to exhibit a weak correlation with ΔvdW. We posit that this is due to the strong affinity of B-O bonds, as indicated by the small bond length of $\sim$1.4 Å in various boronic acids.[47] Such an analysis of the bonding mechanisms is crucial in the selection of organic intercalants to design hybrid materials for specific applications.

In our quest to identify materials suitable for optoelectronic applications, we conducted a comprehensive anal-ysis of the top 50 hybrid materials, focusing specifically on their semiconducting properties. Using the HSE06 hybrid functional, we calculated the electronic structures of the leading hybrid materials and their corresponding parent bilayers. Figure 6 shows the electronic bandgap ($E_g^{HSE}$) of the 40 hybrid materials, ordered by decreasing values, alongside the bandgap of their parent $MX_2$ bilayers. We established that all seven $MX_2$ bilayers contributing to the superior hybrid materials possess semiconducting properties. The bandgap spans from a maximum of 1.88 eV for $HfS_2$ to a minimum of 1.22 eV for $CrO_2$, covering photon energies from the near-infrared (NIR) to the low-energy visible light spectrum. After intercalation, we observed a substantial decrease in $E_g^{HSE}$, with bandgaps ranging from 0.62 eV to as low as 0.07 eV. We attribute these significant variations in bandgap to the strong renormalization of the optoelectronic features by the intercalant, including midgap states. This shift results in an extended absorption spectrum into the infrared region, from NIR to long-wavelength infrared, demonstrating versatile tunability. Such adaptability is crucial for the design of optoelectronic devices that require precise bandgap control, rendering these hybrid materials highly promising for advanced optoelectronic applications.

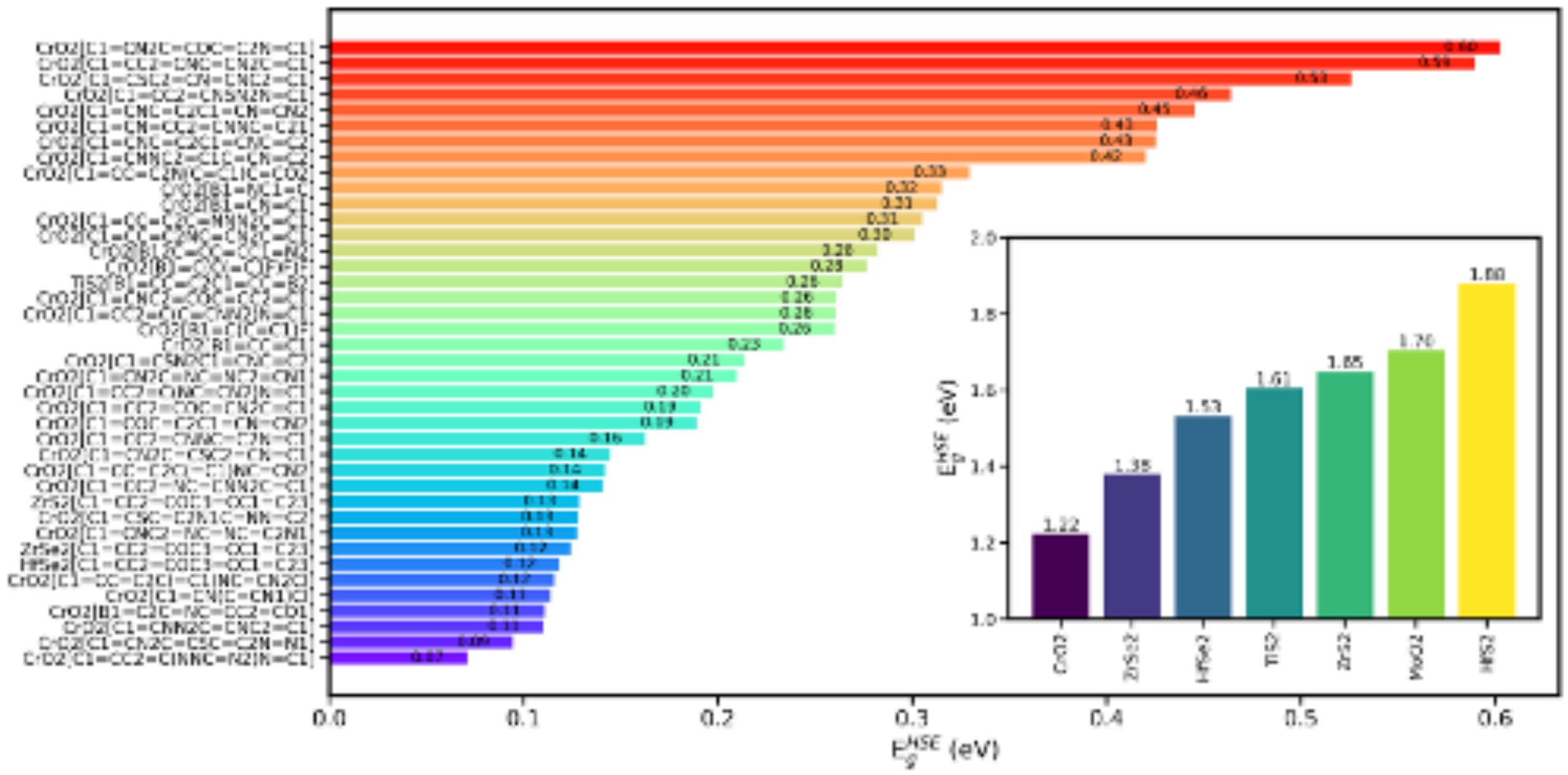

FIG. 6. The HSE06 bandgap ($E_g^{HSE}$) of 40 out of the 50 topmost stable hybrid materials. The inset shows $E_g^{HSE}$ of the corresponding $MX_2$ bilayers.

## IV. Conclusion

We have developed a high-throughput computational approach, combining first-principles methods, materials informatics, and machine learning to design novel quantum materials derived from intercalating conjugated molecules into bilayer transition metal dichalcogenides and dioxide. With our high-throughput, data-driven approach, we identified the 50 most stable hybrid materials from an extensive compositional space comprising over 300,000 combinations, employing intercalation energy as the screening criterion. We identified several hybrid materials with promising optoelectronic features that could be explored for advanced application. The developed data-driven framework is versatile and can easily be extended to a large variety of host materials and intercalants, making it an efficient tool for screening an extensive database of hybrid materials. Additionally, this methodology can be integrated with transfer learning frameworks that leverage low-fidelity DFT band gaps as screening parameters while using high-fidelity experimental or GW band gaps as a reference to design advanced materials with tailored electronic properties.

## V. Supplementary Material

Please refer to the Supplementary Material for a detailed account of the active learning algorithm including model selection (Table S1 and Figure S1) and feature selection (Figure S2). Figures S3-S8 provide additional information on the active learning predictions per iteration, including the stability ranking of the predicted top 50 materials. Tables S2-S4 provide additional information including intercalation energies, mechanical properties, etc.

## Acknowledgments

This research is supported by the U.S. Department of Energy, Office of Science, Basic Energy Sciences under Award DOE-SC0024099 (algorithmic development and first-principles calculations) and the Lehigh Core grant (initial framework development). S.M.K. acknowl-edges the Lee Graduate Fellowship from Department of Physics, Lehigh University. Computational resources were provided by Lehigh University Research Computing Infrastructure through NSF Award 2019035.

### Data availability

The structural data generated for this project, along with the material design pipeline and active learning algorithm codes, are available in the public GitHub repository: https://github.com/gmp007/Custom-Design-of-2D-based-Materials.

### Conflicts of interest

There are no conflicts to declare.

### Notes and references